\documentclass[11pt]{article}
\usepackage{latexsym}
\def\a{\alpha}
\def\b{\beta}

\def\d{\delta}
\def\e{\epsilon}

\def\g{\gamma}

\def\G{\Gamma}

\def\La{\Lambda}
\def\O{\Omega}

\def\U{\Upsilon}

\def\da{{\dot\alpha}} 
\def\db{{\dot\beta}}

\def\ds{{\dot{s}}}

\def\dr{{\dot{r}}}
\def\dv{{\dot{v}}}

\font\sixrm=cmr6
\font\eightrm=cmr8  
\def\sc{\sixrm}

\font\cmssl=cmss10 at 11 pt
\font\cmssll=cmss10 at 12 pt

\def\smallsetminus{\backslash}

\DeclareFontFamily{OT1}{msb}{}{}
\DeclareFontShape{OT1}{msb}{m}{n}
 {  <5> <6> <7> <8> <9> <10> gen * msbm
      <10.95><12><14.4><17.28><20.74><24.88>msbm10}{}
\DeclareMathAlphabet{\bubble}{OT1}{msb}{m}{n}

\def\bR{{\bubble R}}
\def\bZ{{\bubble Z}}
\def\bC{{\bubble C}}
\def\bN{{\bubble N}}
\def\bK{{\bubble K}}

\def\ca{{\cal A}}

\def\cc{{\cal C}}

\def\cg{{\cal G}}

\def\cm{{\cal M}}

\def\cv{{\cal V}}

\def\tableeleven{Table A\ }
\def\tablefourtyeight{Table B\ }
\def\tabletwelve{Table C\ }

\def\uz{$u_{0}$}
\def\umu{$u_{-1}$}
\def\umd{$u_{-2}$}
\def\umt{$u_{-3}$}
\def\vmu{$v_{-1}$}
\def\vmd{$v_{-2}$}
\def\vmt{$v_{-3}$}

\def\wmd{$w_{-2}$}

\def\tmu{$t_{-1}$}

\def\unu{$u_{1}$}
\def\und{$u_{2}$}
\def\unt{$u_{3}$}
\def\unq{$u_{4}$}
\def\vnu{$v_{1}$}

\def\wnu{$w_{1}$}

\def\tnu{$t_{1}$}
\def\tnd{$t_{2}$}
\def\tnt{$t_{3}$}

\def\utu{$\wt u_{1}$}
\def\utd{$\wt u_{2}$}
\def\utt{$\wt u_{3}$}
\def\utq{$\wt u_{4}$}
\def\vtu{$\wt v_{1}$}

\def\wtu{$\wt w_{1}$}

\def\ttu{$\wt t_{1}$}
\def\ttd{$\wt t_{2}$}
\def\ttt{$\wt t_{3}$}

\def\Uz{{\sc U}$_{0}$}
\def\Umd{{\sc U}$_{-2}$}
\def\Vmd{{\sc V}$_{-2}$}
\def\Wmd{{\sc W}$_{-2}$}
\def\Und{{\sc U}$_{2}$}
\def\Utd{$\wt{\mbox{\sc U}}_2$}
\def\Unq{{\sc U}$_{4}$}
\def\Utq{$\wt{\mbox{\sc U}}_4$}
\def\Tnu{{\sc T}$_{1}$}
\def\Tnd{{\sc T}$_{2}$}
\def\Tnt{{\sc T}$_{3}$}
\def\Ttu{$\wt{\mbox{\sc T}}_1$}
\def\Ttd{$\wt{\mbox{\sc T}}_2$}
\def\Ttt{$\wt{\mbox{\sc T}}_3$}
\def\U{{\mbox{\eightrm U}}}
\def\W{{\mbox{\eightrm W}}}

\def\uzm{u_{0}}
\def\umum{u_{-1}}
\def\umdm{u_{-2}}
\def\umtm{u_{-3}}
\def\vmum{v_{-1}}
\def\vmdm{v_{-2}}
\def\vmtm{v_{-3}}

\def\wmdm{w_{-2}}

\def\tmum{t_{-1}}

\def\unum{u_{1}}
\def\undm{u_{2}}
\def\untm{u_{3}}
\def\unqm{u_{4}}
\def\vnum{v_{1}}

\def\wnum{w_{1}}

\def\tnum{t_{1}}
\def\tndm{t_{2}}
\def\tntm{t_{3}}

\def\utum{\wt u_{1}}
\def\utdm{\wt u_{2}}
\def\uttm{\wt u_{3}}
\def\utqm{\wt u_{4}}
\def\vtum{\wt v_{1}}

\def\wtum{\wt w_{1}}

\def\ttum{\wt t_{1}}
\def\ttdm{\wt t_{2}}
\def\tttm{\wt t_{3}}

\def\be{\begin{equation}}
\def\ee{\end{equation}}
\def\bea{\begin{eqnarray}}
\def\la#1{\label{#1}}  
\def\re#1{(\ref{#1})}
\def\arr{\begin{array}{rll}}
\def\ea{\end{array}}
\def\eea{\end{eqnarray}}

\def\op{{\oplus}}

\def\ra{\rightarrow}
\def\wt{\widetilde}
\def\wh{\overline}

\def\fr#1#2{{\textstyle{#1\over\vphantom2\smash{\raise.20ex
        \hbox{$\scriptstyle{#2}$}}}}}                   
\def\half{{\textstyle{1\over 2}}}
\def\3h{\fr32}
\def\M{$\cm$}

\def\sp{super-Poincar\'e algebra}

\begin{document}
\begin{titlepage}
\rightline{hep-th/9804052}
\vskip 2.0 true cm
\begin{center}
{\Large  Lorentz Covariant Spin Two Superspaces}
\vskip 0.7 true cm
{\large Chandrashekar Devchand$^1$\ \ and\ \  Jean Nuyts$^2$}

\vskip 0.6 true cm
{\small
\centerline{devchand@mis.mpg.de, nuyts@umh.ac.be}}
\vskip 0.2 true cm
{\it $^1$ Max-Planck-Institut f\"ur Mathematik in den Naturwissenschaften}\\
{\it Inselstra\ss{}e 22-26, 04103 Leipzig, Germany}\\[5pt]
{\it $^2$  Physique Th\'eorique et Math\'ematique, 
Universit\'e de Mons-Hainaut}\\
{\it 20 Place du Parc, 7000 Mons, Belgium}

\vskip 2.5 true cm
\noindent
{\bf Abstract}
\end{center}
\begin{quote}
Superalgebras including generators having spins up to two and realisable as
tangent vector fields on Lorentz covariant generalised superspaces are
considered.  The latter have a representation content reminiscent of
configuration spaces of (super)gravity theories. The most general canonical
supercommutation relations for the corresponding phase space coordinates allowed
by Lorentz covariance are discussed. By including generators transforming
according to every Lorentz representation having spin up to two, we obtain, 
from the super Jacobi identities, the complete set of quadratic equations for 
the Lorentz covariant structure constants. These defining equations for 
{\it spin two Heisenberg superalgebras} are highly overdetermined. Nevertheless,
non-trivial solutions can indeed be found. By making some simplifying
assumptions, we explicitly construct several classes of these superalgebras.
\end{quote}
\vfill
\end{titlepage}

\section{Introduction}

The super-Poincar\'e algebra is the extension of the Lorentz algebra by the
supersymmetry algebra, a $\bZ_2$-graded extension of the algebra of translation
vector fields by Grassmann-odd Lorentz spinors.  The Grassmann-even subspace 
$\ca_0$ of the supersymmetry algebra contains the spin one translation generator
$X_{\a\db}$ transforming according to the $(\fr12,\fr12)$ Lorentz 
representation and the odd subspace $\ca_1$ contains the spin $\fr12$ 
representations, $X_{\dot\alpha}$ and $X_{\alpha}$, transforming as 
$(0,\fr12)$ and $(\fr12,0)$. They satisfy
\be
\left\{  X_{\a} , X_{\db}\right\}  =\  2i\  X_{\a\db}\  ,
\label{st1} 
\ee
with all other supercommutators equal to zero. 

The possibility of extending the supersymmetry algebra to include generators 
of spin greater than one, and thus going beyond the Haag-\L{}opusanski-Sohnius
barrier, was broached by Fradkin and Vasiliev \cite{FV,V}. These
authors were motivated by physical considerations to realise such higher-spin
algebras on de Sitter space fields.  Consistency of the dynamics required the
inclusion of {\it all} spins, yielding infinite dimensional algebras realised 
on infinite chains of fields having spins all the way up to infinity.

The approach we have taken recently \cite{DN1,DN2} has been more abstract. 
We considered extentions of the supersymmetry algebra by further representations
of the Lorentz group to those given above, maintaining the $\bZ_2$-grading, 
with all integer-spin representations in the even-statistics (bosonic) subspace
$\ca_0$ and all half-integer-spin representations in the odd-statistics
(fermionic) subspace $\ca_1$.  Insisting on Lorentz covariance determines the
space of the a priori allowed
structure constants and solutions of the super Jacobi identities then
provide concrete examples of Lorentz covariant generalisations of the \sp. From
the work of Fradkin and Vasiliev, it is clear that finite-dimensional examples 
of these algebras do not have non-trivial realisations on fields in standard
four-dimensional space. In \cite{DN1,DN2}, however, the possibility was raised
of realising these algebras on higher-dimensional extensions of four-dimensional
space, generalising the idea of superspace. These {\it hyperspaces} provide
natural representation spaces for finite-dimensional examples of higher-spin
superalgebras.

Recall that the \sp\ is realisable as a superalgebra of infinitesimal
translation vector fields on superspace, the quotient of the super Poincar\'e
group by the Lorentz group, with supercommuting coordinates 
$\ \{Y^\a, Y^\da, Y^{\a\da}\}\,$, having the same Lorentz-transformation 
properties and statistics as the corresponding supersymmetry generators. 
The action of the supersymmetry algebra on superfields depending on these 
coordinates is determined by the Heisenberg superalgebra with non-zero 
canonical supercommutation relations,
\be
\arr
\left[  X_{\a\da}, Y^{\b\db} \right]    =  \d^\b_\a  \d^\db_\da\  ,\quad
\left\{  X_{\a} , Y^{\b } \right\}   =   \d^\b_\a\  ,\quad
\left\{  X_{\da}, Y^\db  \right\}   =  \d^\db_\da \\[5pt]
\left[  X_{\a} , Y^{\b\db} \right]    = i\  \d^\b_\a Y^\db \  ,\qquad
\left[  X_{\da}, Y^{\b\db}\right]   =  i  \d^\db_\da Y^{\b}\  \  .
\la{st2}
\ea
\ee
It is this construction which we generalised in \cite{DN1,DN2}. In these papers
explicit examples of Lorentz covariant hyperspaces \M\ with sets of coordinates 
$\{Y(s,\ds)\}$ 
transforming according to more general $(s,\ds)$ representations than  
the traditional spinorial and vectorial ones were presented. Such hyperspaces 
are graded vector spaces and the coordinates $\{Y(s,\ds)\}$ span a 
supercommutative $\bZ_2$-graded algebra, $\,\cv{=}\cv_0 {+}\cv_1\,$, with 
$\cv_0 $ (resp. $\cv_1$) containing  bosonic (resp. fermionic) coordinates with 
$2(s{+}\ds)$ even (resp. odd). Each set of coordinates $\ Y(s,\ds)\ $ 
included increases the bosonic (resp. fermionic) dimension of \M\ by 
$(2s{+}1)(2\ds{+}1)$. If $p$ is the maximum value of $\,s{+}\ds\,$
occurring, we call the hyperspace a {\it spin $p$ superspace}. 

Tangent spaces of these hyperspaces \M\ are $\bZ_2$-graded vector spaces
$\ca{=}\ca_0{+}\ca_1$ spanned by infinitesimal generalised translation vector 
fields $\{X(s,\ds)\}$.
These vector spaces are required to be superalgebras
(generalising the supersymmetry algebra); i.e.  they have a supersymmetric
bilinear map (super commutator), $\ [\ .\, ,\, .\, ] :  \ca \times\ca\ra\ca\ $,
satisfying the super Jacobi identities.

The vector fields $X\in \ca$ act as superderivations on functions of the $Y$'s.
We assume that the action of $\ca$ on $\cv$ corresponds to a linear
transformation; the combined vector space $\ \cg{=}\ca{+}\cv\ $ having the
supercommutation relations of a generalised Heisenberg superalgebra, 
\be\arr
\ca\times\ca &\ni\ (X,X') \mapsto &[X,X']\, \in\  \ca \\
\ca\times\cv &\ni\ (X,Y)\  \mapsto &[X,Y]\ \in\  \cv\ +\ \cc\\
\cv\times\cv &\ni\ (Y,Y')\, \mapsto &[Y,Y']\ =\ 0\ .
\la{alg}\ea\ee 
Here, $\cc$ is a space of central charges determined by a pairing between $\ca$
and $\cv$ (see section 2.1). The super Jacobi identities are satisfied and the
combined grading is preserved, i.e.
\be
[\ca_\a ,\ca_\b] \subset \ca_{\a+\b}\ ,\quad
[\ca_\a ,\cv_\b] \subset \cv_{\a+\b} + \cc \d_{\a,\b}\  ,
\mbox{ with } \a,\b\in \bZ_2\ .
\ee
We call the algebra $\cg$ with relations \re{alg} a {\it spin p Heisenberg 
superalgebra} if $p$ is the maximum value of $\,s{+}\ds\,$ amongst the 
representations appearing in $\,\ca{+}\cv\,$. 

If the elements appearing in the supercommutators in \re{alg} transform
respectively as $(s,\ds)$ and $(r,\dr)$ representations, Lorentz covariance
requires that the {\it a priori} elements on the right-hand-sides transform 
according to $(v,\dot v)$ representations occurring in the double Clebsch-Gordon
decomposition of the direct product of the two Lorentz representations. Namely,
\bea
s\otimes r &=&\sum \op\ v\ =\ (s{+}r)\,\op\,(s{+}r{-}1)\,\op\dots\op\,|s-r|  
\nonumber \\[4pt]
\ds\otimes\dr&=&\sum \op\ \dv\ =\ 
            (\ds{+}\dr)\,\op\,(\ds{+}\dr{-}1)\,\op\dots\op\,|\ds-\dr|\quad. 
\la{clebschseries}\eea
With this algebraic structure, the hyperspaces \M\ provide higher
dimensional spaces having manifest four-dimensional Lorentz covariance.  They
are modeled on standard superspace used in supersymmetric field theories.
Explicit examples of algebras $\cg$ were presented in \cite{DN1,DN2} for spins
$s{+}\ds$ up to $\fr32$.  As an application gauge fields on \M\ were
considered:  Associating a gauge potential $A$ to each of these generalised
derivatives, we defined, in a natural fashion, the covariant derivative
$\ {\cal{D}}{=}X{+}A\ $ and the corresponding curvature tensors $F$.  This 
allowed us to define generalised self-dualities in terms of Lorentz covariant 
constraints on components of the curvature.  We thus obtained a different class
of higher-dimensional generalisations of the self-duality equations to those
presented in \cite{CDFN}, having manifest four-dimensional Lorentz covariance
and affording generalised twistor-like transforms.  Moreover, a novel hierarchy
of {\it light-like integrable systems} was also presented, whose simplest
non-trivial member is the well-known $N{=}3$ super-Yang-Mills set of on-shell
curvature constraints.  These systems therefore provide an infinitely large
hierarchy of gauge- and Lorentz-covariant solvable systems.

The purpose of this paper is to highlight another setting for the application 
of these higher-spin algebras.  Our hyperspaces \M\ in fact serve as models for
configuration spaces, or for moduli spaces of solutions, of Lorentz invariant
field theories; and the supercommutations relations for $\cg$ provide canonical
supercommutation relations for the corresponding phase spaces, providing an 
algebraic description of the local symplectic structure. With this application
in mind, it is clear that algebras including spins up to two are of possible 
relevance for the canonical quantisation of gravity and supergravity theories. 
In \cite{DN1,DN2}, a rather simple non-trivial example containing generators of
spins $\fr12, 1$ and $\fr32$ was presented.  In order to search for possibly
interesting examples containing spin $2$ generators, we use a modified notation
to that of \cite{DN1,DN2}, which is more convenient for the extraction of the
complete set of algebraic equations for the structure constants from the 
super Jacobi identities.  We describe these in section 2.  In section 3,
we restrict ourselves to superalgebras $\cg$ containing all Lorentz tensors
having spin less than or equal to $2$.  For unit multiplicity of each Lorentz
representation $(s,\ds)$ for $0\le s{+}\ds\le 2\ $, we obtain, for the
superalgebra $\ca\ $, $\,1993\ $  quadratic equations for $\ 163\ $
structure constants, which are supplemented by further $\ 5732\ $ quadratic 
equations for a total of $\ 163 + 339\ $ structure constants and $15$ central 
charges. Each solution of this overdetermined system of equations corresponds
to a specific example of a spin~$2$ Heisenberg superalgebra $\cg\ $; and we 
present some classes of solutions in section~4.
In fact, the space of $\ 163{+}339{+}15{=}517\ $ structure constants subject to 
$\ 1993{+}5732{=}7725\ $ quadratic equations parametrises the moduli space of
spin~$2$ Heisenberg superalgebras.

Remarks:

\noindent
a) Our setting is basically complex: We consider representations of the complex
extension of the Lorentz algebra, $so(4,\bC) = sl(2,\bC)\op sl(2,\bC)$. 
This yields higher spin superalgebras with or without the `chiral' symmetry 
interchanging dotted and undotted indices, which for the real Lorentzian case
is an automatic consequence of complex conjugation. The broader complex 
framework thus affords more general possibilities, which may be of relevance
in concrete physical settings requiring covariance under a Euclidean 
($\ so(4)=su(2)\op su(2)\ $), Lorentzian ($\ so(3,1)=sl(2,\bC)\ $) or Kleinian 
($\ sl(2,\bR)\op sl(2,\bR)\ $) real six-dimensional subalgebras of 
$\ so(4,\bC)$.

\noindent
b) Although we remain in the realm of supercommutative geometry, with
$\ [\cv,\cv]{=}0\,$, a generalisation to non-supercommutative geometry is 
clearly a further possibility, with the simplest superalgebra variant 
having $\ [\,.\,,\,.\,] : \cv \times \cv \ra \cv\ $ such that 
$\ [\cv_\a ,\cv_\b] \subset \cv_{\a+\b}\ $. Further generalisations, replacing
this superalgebra structure, for instance, by $q$-deformed supercommutation 
relations, may also be considered along the lines of the present investigation. 

\noindent
c) In this paper, we will consider an element of $\ca, \cv$ to be of bosonic 
type if its spin $(s{+}\ds)$ is an integer and of fermionic type if its spin is
a genuine half-integer; and we use shall assume the corresponding statistics.
We note, however, that the assignment of even (resp. odd) statistics to elements
of $\ca_0, \cv_0$ (resp. $\ca_1, \cv_1$) is a purely conventional one, motivated
by the spin-statistics theorem. This can indeed be lifted, if required, to yield
Lie algebra (rather than superalgebra) extensions of the Poincar\'e algebra 
containing integer and half-integer spin elements, all of even statistics. Such 
algebras maintain, nevertheless, their $\bZ_2$-graded nature \cite{AC}. 
Such a variant of the supersymmetry algebra was recently shown to be the target
space symmetry of the $N{=}2$ string \cite{DL} and the space of string physical
states was shown to be elegantly and compactly describable in terms of a field 
on a hyperspace with a vectorial and an even-spinorial coordinate.

\noindent
d) In the context of application of our formalism to canonical quantisation of 
(super) gravity theories, we have only considered the simplest putative phase
space coordinates: a metric represented by canonically conjugate variables
$X,Y$ transforming as $(0,0)+(1,1)$  coupled to single copies of other 
representations.
A generalisation to higher multiplicities ($N>1$ supergravities) follows on 
the lines of the discussion in appendix A of \cite{DN2}, with the variables
acquiring a further `internal' index labeling the inequivalent copies of any 
particular representation thus:\ 
$\{X(s,\ds)\}\ ,\ \{Y(s,\ds)\}\ra 
\{X(s,\ds\,;n)\}\ ,\ \{Y(s,\ds\,;m)\}$. 

\noindent
e) A further interesting generalisation is to matrix-indexed variables $X,Y$.
This is clearly a variant of the above-mentioned higher-multiplicity
generalisation, with the variables having additional indices labeling a space 
of internal matrices rather than internal vectors. For instance, Ashtekar's
canonical variables for gravity consist of $\ sl(2,\bC)$-indexed
$X(1,0),Y(1,0)$.

\section{Higher-spin superalgebras}
In this section we give a more precise definition of
{\it spin p Heisenberg superalgebras} $\cg$. For simplicity, we restrict 
ourselves to the case of unit multiplicity of any tensor with given Lorentz 
behaviour.  A generalisation to `N-extended' cases is, in principle, 
straightforward; based on the discussion in the Appendix of \cite{DN2}.

Let us denote by $\Lambda_p$ some freely specifiable lattice of 
doublets of half-integers $(s,\ds)$,
\be
\La_p = \{ (s,\ds) \}\ \subset \bK^2 \quad,\quad 
{\mbox{with}}\quad p={\mbox{max}}\{s{+}\ds\} \quad ,    
\la{lattice}
\ee 
where $\bK=\fr12\bN\cup\{0\}\ $, the set of non-negative half-integers.

\subsection{A basis for $\cg{=}\ca{+}\cv$}

For any point $(s,\ds)$ in $\Lambda_p$, consider the coordinate tensor 
$\,Y(s,\ds)$ transforming according to the $(s,\ds)$ representation of 
the Lorentz group.  We label its $(2s{+}1)(2\ds{+}1)$ components as 
$\,Y(s,s_3\,;\ds,\ds_3) $, where $s_3$ (resp.  $\ds_3$) run from
$-s$ to $s$ (resp.  from $-\ds$ to $\ds$) in integer steps. 

We define the span of coordinates $\{ Y(s,s_3\,;\ds,\ds_3)\}$, 
for all $(s,\ds)$ in the chosen set $\Lambda_p$, to be a basis of the vector 
space $\cv$. This provides a coordinate system for Lorentz-covariant 
spin $p$ hyperspaces \M.  
The corresponding tangent space $\ca\ $ is spanned by the components
$\{ X(s,s_3\,;\ds,\ds_3)\} $ of vector fields $\{X(s,\ds)\}$. 
These vector fields are taken to be in one-to-one correspondence with the 
coordinate tensors, and therefore with the
points on the lattice $\La_p$. The specific choice of points making up this
lattice therefore effectively determines
the basis elements of both vector spaces $\ca$ and $\cv$. 
So, for instance, the superalgebra \re{st1},\re{st2} is based on the lattice
of three points $\ \La_p=\{(0,\fr12),(\fr12,0),(\fr12,\fr12)\}$. Between 
vector field $X(s,\ds)$ and coordinate tensor $\,Y(s,\ds)$, which transform 
similarly under the Lorentz group, we assume a bilinear pairing given by
\be\arr 
&& <X(s,s_3\,;\ds,\ds_3), Y(r,r_3\,;\dr,\dot r_3)>\\[4pt]
&&\quad =\ 
c(s,\ds)\
C(s,s_3,s,-s_3\,;0,0)\
           C(\ds,\ds_3,\ds,-\ds_3\,;0,0)\
 \delta_{sr}\,\delta_{\ds\dr}\, \delta_{s_3+r_3,0}\,
\delta_{\ds_3+\dot r_3,0}\quad . 
\ea\la{pair}\ee
Here the Clebsch-Gordon coefficients $C(s,s_3,s,-s_3\,;0,0)\ $ denote Wigner's 
`metric' invariant in the representation space of fixed spin $s$ 
(see e.g. \cite{Var}). This defines a pairing map 
\be
 <.\,,\,.>\, :\, \ca \times \cv 
\ra \cc\ =\ \{ c(s,\ds)\ ;\ (s,\ds)\in\La_p \}\  .
\ee
The coefficients $c(s,\ds)$, which will be called the 
central structure constants, can, in principle, be zero or, if 
non-zero, can be set to $1$ by a suitable renormalisation of the $X$'s 
and/or the $Y$'s (provided, as is the case here,
that representations do not occur multiply).
Thus, $c\,:\,\La_p \ra \bZ_2 = \{0,1\}\,$. We shall henceforth, without loss of 
generality, assume that the $c$'s are thus renormalised.

The above way of representing the $(2s+1)(2\ds +1)$ components of an 
$(s,\ds)$-tensor is equivalent to the representation in standard two-spinor
index notation with $2s$ (resp. $2\ds$) symmetrised undotted (resp. dotted)
indices, e.g.  $X_{\a\dots\a_{2s}}^{\,\,\,\da\dots\da_{2\ds}}\ $.  The
two-spinor notation, which was used in \cite{DN1,DN2}, has the advantage of
having the (double) Clebsch-Gordon decomposition readily expressible in terms 
of products of the invariant two-index $\e$-tensors, viz.  $\e^{\a\b}$ and
$\e_{\da\db}$.  Obtaining the complete set of quadratic defining
conditions for the structure constants, however, is not a very straightforward
procedure.  In the above alternative non-index notation, the quadratic equations
may be found using a purely algorithmic procedure using the super
Jacobi identities and explicit values for Clebsch-Gordan coefficients and $6j$
symbols from e.g.  \cite{Var}.  The extraction of {\it all} the quadratic 
conditions for the structure constants is then streamlined, allowing automation
of the procedure using a symbolic manipulation language like {\small REDUCE}
or {\small MAPLE}.  Once a solution of these equations
is found, the supercommutation relations for the algebra $\cg$ may be written
immediately in either notation.  The correspondence between components in the
two notations may easily be established.  For instance, the index-notation
component of $X(s,\ds)$ with $n$ (resp.  $\dot n$) indices taking the value
$1$ (resp.  $\dot 1$), with the remaining $2s{-}n$ undotted (resp.  
$2\ds{-}\dot n$ dotted) indices taking the value $2$ (resp.  $\dot 2$), 
denoted $X(1^n 2^{(2s{-}n)} \dot 1^{\dot n} \dot 2^{(2\ds{-}\dot n)})$, is 
related to the $s_3{=}n{-}s\ ,\ \ds_3{=}\dot n{-}\ds\ $ component thus:
\be
X(1^n 2^{(2s{-}n)} \dot 1^{\dot n} \dot 2^{(2\ds{-}\dot n)}) =
P(s,\ds) \sqrt{\left( n! (2s-n)! \dot n! (2\ds-\dot n)! \right)}
X(s,n-s ;\ds,\ds-\dot n)\ ,
\ee
where $P(s,\ds)$ is an arbitrary normalisation.

\subsection{The supercommutation relations}
We take the entire set of coordinates $\{Y(s,s_3\,;\ds,\ds_3)\}$, 
for all $(s,\ds)\in \Lambda_p$, to be supercommutative:
\begin{equation}
\left [\ Y(s,s_3\,;\ds,\ds_3)\ ,\
Y(r,r_3\,;\dr,\dot r_3)\ \right ]_{S\bullet R}=0\ , 
\label{comYY}
\end{equation}
where we introduce the shorthand notation
\be
S=(s,\ds)\  ,\quad R=(r,\dr)\ ,
\label{labelL2}
\ee
in terms of which the sign of the graded bracket is defined as
\begin{equation}
S\bullet R=R\bullet S =(-1)^{4(s+\ds)(r+\dr)+1}\  .
\label{SbulletT} 
\end{equation}

We define another lattice in two dimensions, $\G(S,R)\subset \bK^2$,
to be the set of representation labels $(v,\dv)$ arising in the Clebsch-Gordon
product \re{clebschseries} of $(s,\ds)$ with $(r,\dot{r})$, namely, 
\be
\G(S,R) = \{\ (v,\dv)\ ;\quad v \in \g(s,r)\ ,\   
\ \dot v \in\ \g(\ds,\dr) \}\ .
\la{gamma}
\ee
Here we denote by $\g(s,r)\subset\bK$ the set of integers
or half-integers arising in any single Clebsch-Gordon series,
\be
\g(s,r)\ =\ \{\ s{+}r\,,\ s{+}r{-}1\,,\ \dots,\ |s-r|\ \}\ ,
\ee 
so that the lattice in \re{gamma} $\G(S,R)=\g(s,r)\otimes\g(\ds,\dr)$.

We postulate that the vector fields $\{X(s,s_3;\ds,\ds_3)\}$ generate a 
Lorentz covariant superalgebra $\ca\ $. The most general supercommutation 
relations allowed by Lorentz covariance have the form 
\bea
&& \left [\ X(s,s_3\,;\ds,\ds_3)\ ,\
     X(r,r_3\,;\dr,\dot r_3)\ \right ]_{S\bullet R}  
\nonumber \\[8pt]
     &&=\quad \sum_{(v,\dv)\in\G(S,R)\cap\La_p}
     C(s,s_3,r,r_3\,;v,s_3+r_3)\ 
     C(\ds,\ds_3,\dr,\dot r_3\,;\dv,\ds_3+\dot r_3)
\nonumber \\
     &&\hspace{3.5truecm}\times\, t(s,\ds\,,r,\dr\,,v,\dv)\ 
     X(v,s_3+r_3\,;\dv,\ds_3+\dot r_3)\quad .
\label{comXX}
\eea 
Here the Clebsch-Gordan coefficients $C(s,s_3,r,r_3\,;v,s_3+r_3)$ 
have the symmetry property
\be
C(s,s_3,r,r_3\,;v,s_3+r_3)\ =\ (-1)^{s+r-v}C(r,r_3,s,s_3\,;v,s_3+r_3)\quad.
\label{CGsym}
\ee
The super Jacobi identities for the supercommutation relations \re{comXX}
yield quadratic equations for the set of admissible structure constants 
$t(s,\ds\,,r,\dr\,,v,\dv)$. Solutions then define superalgebras $\ca$. 
For any choice of $\La_p$ the admissible structure constants depend on 
six spin variables, integers or half-integers specifying a lattice of points
in six dimensions, $\ \O_p\subset \bK^6\ $, defined by
\be
\O_p\ =\ \{\ (s,\ds\,,r,\dr\,,v,\dv)\ ;\quad 
(s,\ds), (r,\dr) \in \La_p\ ,\
(v, \dv)\in \G(S,R) \cap  \La_p\ \}
\la{omega}
\ee
The space of these structure constants is 
manifestly restricted by superskewsymmetry, namely,
\be
t(r,\dr\,,s,\ds\,,v,\dv)\ =\
(-1)^{4(s+\ds)(r+\dr)+(s+\ds)+(r+\dr)-(v+\dv) +1}\,
t(s,\ds\,,r,\dr\,,v,\dv)\quad .
\label{skewsym}
\ee
This redundancy in the set of structure constants may be factored out with no 
loss of generality by imposing the restriction 
$\Omega_p \mid_{S\leq R}$, where the ordering $\,S\leq R\,$ denotes 
$\,s{+}\ds\leq r{+}\dot{r}\ $ and for equality, $s\leq r\ $.  
Equation \re{skewsym} also implies that certain parameters vanish, viz.,
\be 
t(s,\ds\,,s,\ds\,,v,\dv) = 0 \quad\mbox{if}
\quad 4(s{+}\ds)^2{+}2(s{+}\ds){-}(v{+}\dv){+}1 = 1\ \mbox{mod}\ 2\ ,
\la{zeroes}\ee
i.e. if both $\ 2(s{+}\ds)\ $ and $\ 2(s{+}\ds)-(v{+}\dv)\ $ are either even 
or odd.

We require that the vector space $\cv$ carries a linear representation of 
$\ca\,$. This then allows realisation of this superalgebra by vector fields 
satisfying \re{comXX} and acting as superderivations on functions of the $Y$'s.
The most general Lorentz covariant supercommutation relations between the $X$'s
and the $Y$'s consistent with this requirement take the form
\bea
&&\left [\ X(s,s_3\,;\ds,\ds_3),
Y(r,r_3\,;\dr,\dot r_3)\ \right ]_{S\bullet R}  
\nonumber \\[8pt]
&&=\quad \sum_{(v,\dv)\in\G(S,R)\cap\La_p}
\quad    C(s,s_3,r,r_3\,;v,s_3+r_3)\,
           C(\ds,\ds_3,\dr,\dot r_3\,;\dv,\ds_3+\dot r_3)
\nonumber \\
 &&\hspace{3.9truecm} \times\,      u(s,\ds\,,r,\dr\,,v,\dv)\
     Y(v,s_3+r_3\,;\dv,\ds_3+\dot r_3)
\nonumber \\[10pt]
&&\quad +\quad   C(s,s_3,s,-s_3\,;0,0)\
           C(\ds,\ds_3,\ds,-\ds_3\,;0,0)\
      c(s,\ds)\ \delta_{sr}\,\delta_{\ds\dr}\,
      \delta_{s_3+r_3,0}\, \delta_{\ds_3+\dot r_3,0}
\ \ \ .
\label{comXY}
\eea 
The $u$'s are further structure constants, also depending on the lattice
$\Omega_p\,$, i.e. $u :\,\Omega_p \rightarrow \bC$. They have no a priori 
symmetry properties under interchange of points on $\Omega_p\ $. The $X$'s thus 
transform the $Y$'s linearly amongst themselves and the combined vector space 
$\,\cg{=}\ca{+}\cv\,$ forms an enlarged superalgebra if the structure 
constants $\{t,u,c\}$ are subject to the quadratic equations tantamount to the 
satisfaction of the super Jacobi identities amongst the $X$'s and the $Y$'s.

\subsection{The super Jacobi identities}
Refining the shorthand notation \re{labelL2},
\be 
\overline{S}=\{s,s_3\,;\ds,\ds_3\}\ ,\quad
\overline{R}=\{r,r_3\,;\dr,\dot r_3\}\ ,\quad\mbox{etc.},
\label{lableL1}
\ee
the super Jacobi identities for any three operators $A(\overline S)$,
$B(\overline R)$ and $C(\overline V)$ are given by
\bea
&&\left [ \left [ A(\overline S),B(\overline R)\right ]_{S\bullet R},
C(\overline V)\right ]_{(S+R)\bullet V}
\nonumber \\[8pt]
&&\quad - (S\bullet R)
\left [  B(\overline R),\left [A(\overline S),
C(\overline V)\right ]_{S\bullet V}\right ]_{(S+V)\bullet R}
\nonumber \\[8pt]
&&\quad - \left [  A(\overline S),\left [B(\overline R),
C(\overline V)\right ]_{R\bullet V}\right ]_{(R+V)\bullet S}
\quad =\quad 0
\label{superjacobi}
\eea
Since the $Y$'s supercommute (\ref{comYY}), the only non trivial 
(not automatically satisfied) super Jacobi identities are the ones for three 
$X$'s and for two $X$'s and a $Y$.

For any particular choice of $\overline S,\overline R,\overline V$ the identity 
\re{superjacobi} yields, in general, several equations for the structure 
constants $t,u$ and $c$, since  the coefficients of all the linearly
independent tensors have to vanish. These can be determined by
using the recoupling or $6j$-symbols. In particular, we require the
formula (see (\cite{Var}, eq.(36), p.261)) 
\bea
&&C(b,b_3,a,a_3\,;e,e_3)\,C(c,c_3,e,e_3\,;f,f_3)
\label{6jdef}\\[8pt]
&&=\sum_{s,s_3}(-1)^{2f}\left\{ \matrix{a&b&e\cr
                                     c&f&s   } \right\}
(2e+1)^{1\over 2}(2s+1)^{1\over 2}  
\, C(b,b_3,c,c_3\,;s,s_3)\,C(a,a_3,s,s_3\,;f,f_3)
\nonumber 
\eea
where the indices are restricted to their obvious allowed ranges, viz.,
on the left side, 
\be\arr
&e\ \in \g(a,b)\quad,\quad  & e_3=a_3+b_3 \\
&f\ \in \g(c,e)\quad,\quad  &f_3=c_3+e_3=a_3+b_3+c_3 
\ea\ee
specifying the domain of definition of the $6j$-symbols; and on the 
right side, determining the ranges of the $s,s_3$ summation,
\be
s\in \g(b,c)\cap\g(f,a)\quad,\quad  s_3=b_3+c_3 \quad .
\ee

The super Jacobi identities between $X(\overline S)$, $X(\overline R)$ and 
$X(\overline V)$ yields quadratic equations for the structure constants $t$
in \re{comXX}. Using \re{comXX} and \re{6jdef} we obtain   
\\ {\underline{\it the tt-equations:}}
\bea
&& t(s,\ds\,,r,\dr\,,e,\dot e)\,t(e,\dot e\,,v,\dv\,,f,\dot f) 
\nonumber \\[10pt]
&& -\quad\sum_{g,\dot g}(S\bullet R) 
         (-1)^{e+v+f+\dot e+\dv+\dot f}
         \sqrt{(1+2g)(1+2e)(1+2\dot g)(1+2\dot e)}
\nonumber \\
&&\quad \quad\quad\times\left\{ \matrix{v&s&g\cr
                                        r&f&e    }\right\}
\left\{ \matrix{\dv&\ds&\dot g\cr
                \dr&\dot f&\dot e   }\right\}
t(s,\ds\,,v,\dv\,,g,\dot g)\,t(r,\dr\,,g,\dot g\,,f,\dot f) 
\nonumber \\[10pt]
&&-\quad \sum_{h,\dot h} (-1)^{s+r+v+f+\ds+\dr+\dv+\dot f}
\sqrt{(1+2h)(1+2e)(1+2\dot h)(1+2\dot e)}
\nonumber \\
&&\quad\quad\quad\times\left\{\matrix{v&r&h\cr
                                      s&f&e}\right\}
\left\{\matrix{\dv&\dr&\dot h\cr
              \ds&\dot f&\dot e}\right\}
t(r,\dr\,,v,\dv\,,h,\dot h)\,t(s,\ds\,,h,\dot h\,,f,\dot f)
\quad=\quad 0\ . 
\label{ttequations}
\eea \\[4pt]
These equations are to be imposed for every $\ S,R,V\in\La_p$, corresponding to
the three operators appearing in the super-Jacobi identities \re{superjacobi}, 
and for every possible intermediate and final-state indices, viz.,
\be\arr
(e,\dot e)=E&\in& \G(S,R)\cap\La_p \\[5pt]
(f,\dot f)=F&\in& \G(E,V)\cap\La_p \quad.
\label{EFrange}
\ea\ee
The ranges of the summations in \re{ttequations} are given by
\be\arr
(g,\dot g)=G&\in& \G(S,V)\cap\G(R,F)\cap\La_p \\[5pt]
(h,\dot h)=H&\in& \G(R,V)\cap\G(S,F)\cap\La_p \quad.
\ea\label{GHrange}
\ee
Interchanging the indices $S$, $R$ and $V$ clearly does not produce independent
equations, so that these indices need to be restricted by some ordering, e.g.
$S\leq R\leq V$.
The space of parameters $t(s,\ds\,,r,\dr\,,v,\dv)$, with domain given by 
\re{omega} and subject to \re{skewsym} and \re{ttequations} is the parameter
space of superalgebras $\ca\,$. 

\goodbreak
The super Jacobi identities between operators $X(\overline S), 
X(\overline R)$ and $Y(\overline V)$  yield
\\ {\underline{\it the tu-equations:}}
\bea
&& t(s,\ds\,,r,\dr\,,e,\dot e)\,u(e,\dot e\,,v,\dv\,,f,\dot f) 
\nonumber \\[10pt]
&& - \sum_{g,\dot g}(S\bullet R) 
    (-1)^{e+v+f+\dot e +\dv+\dot f}
    \sqrt{(1+2g)(1+2e)(1+2\dot g)(1+2\dot e)}
\nonumber \\[5pt]
&&\quad\quad\times\left\{ \matrix{v&s&g\cr
                r&f&e    }\right\}
\left\{ \matrix{\dv&\ds&\dot g\cr
                \dr&\dot f&\dot e   }\right\}
u(s,\ds\,,v,\dv\,,g,\dot g)\,u(r,\dr\,,g,\dot g\,,f,\dot f) 
\nonumber \\[10pt]
&& - \sum_{h,\dot h} 
      (-1)^{s+r+v+f+\ds+\dr+\dv+\dot f}
      \sqrt{(1+2h)(1+2e)(1+2\dot h)(1+2\dot e)}
\nonumber \\[5pt]
&&\quad\quad\times\left\{\matrix{v&r&h\cr
              s&f&e}\right\}
\left\{\matrix{\dv&\dr&\dot h\cr
              \ds&\dot f&\dot e}\right\}
u(r,\dr\,,v,\dv\,,h,\dot h)\,u(s,\ds\,,h,\dot h\,,f,\dot f)
\quad =\quad 0 \quad.
\label{tuequations}
\eea \\[4pt]
These equations hold for every $\ S,R,V\in\La_p\ $, with an ordering $S\leq R$, 
and every allowed $E,F$ given in (\ref{EFrange}). The summations are again over
values of $G,H$ in \re{GHrange}.
The $(X(\overline S)$, $X(\overline R)$, $Y(\overline V))$-identities also yield
\\ {\underline{\it  the tuc-equations:}}
\bea
&& t(s,\ds\,,r,\dr\,,v,\dv)\,c(v,\dv) 
\nonumber \\[10pt]
&&\quad - (S\bullet R) (-1)^{2r+2\dv}
\sqrt{(2r+1)(2r+1)(2\dr+1)(2\dv +1)}
\nonumber \\[5pt]
&&\quad\quad\times\left\{\matrix{v&s&r\cr
                                 r&0&v}\right\}
\left\{\matrix{\dv&\ds&\dr\cr
                     \dr& 0&\dv}\right\}
u(s,\ds\,,v,\dv\,,r,\dr)\,c(r,\dr) 
\nonumber \\[10pt]
&&\quad - (-1)^{s+r+v+\ds+\dr+\dv}
\sqrt{(2s+1)(2r+1)(2\ds+1)(2\dv +1)}
\nonumber \\[5pt]
&&\quad\quad\times\left\{\matrix{v&r&s\cr
                                 s&0&v}\right\}
\left\{\matrix{\dv&\dr&\ds\cr
                     \ds& 0&\dv}\right\}
u(r,\dr\,,v,\dv\,,s,\ds)\,c(s,\ds) \quad =\quad 0 
\label{tcequations}
\eea \\[4pt]
for every $\ S,R,V\in\La_p\ $.  Again, an ordering $S\leq R$ yields independent
equations, which are in one-to-one correspondence with the number of 
inequivalent non-zero $t$'s.

The space of parameters $t(s,\ds\,,r,\dr\,,v,\dv)$ and
$u(s,\ds\,,r,\dr\,,v,\dv)$, with domain given by $\O_p$ and the t-equivalences
\re{skewsym} and zeroes \re{zeroes} factored out, together with the central 
charges $c(s,\ds)$, with $S\in \La_p$, subject to the quadratic constraints 
\re{ttequations},\re{tuequations} and \re{tcequations}, is the moduli space of 
spin $p$ Heisenberg superalgebras.
Particular solutions of the equations \re{ttequations},\re{tuequations} and 
\re{tcequations} correspond to examples of superalgebras $\cg$.
In section 4, we construct some explicit classes of solutions for
values of spin up to 2, i.e. for various choices of $\La_2\ $.

\section{Superalgebras $\cg$ including elements of spin up to two}

With the values of $6j$-symbols taken from the tables of \cite{Var}, we have 
used {\small REDUCE} to
generate the complete set of quadratic equations for the structure constants 
$\{t,u,c\}$ of superalgebras $\cg$ containing elements 
$\{X(s,\ds),Y(s,\ds)\}$ for spins up to $s{+}\ds =2$, i.e. with
generators having indices spanning the largest $\La_2$ lattice,
\be \La_2^{\sc{max}} = 
\{(s,\ds)\,;\ s,\ds \in \bK\ ,\ 0 \leq s{+}\ds \leq 2\}\ .
\la{spin2set}\ee
Specifically, the representations we have taken into account,
with multiplicity one, are

\begin{enumerate}
\item  spin $0$ 
  \begin{itemize}
  \item a scalar $(0,0)$. The variable $Y(0,0)$ could correspond to the trace 
   of the space time metric $g_{\mu\nu}$, representing the conformal weight. 
   The corresponding vector field $X(0,0)$ generally behaves like a dilatation 
   operator, generating Weyl rescalings of the metric.
  \end{itemize}
\item  spin $\fr12$
  \begin{itemize}
  \item a Dirac spinor $(\fr12,0)+(0,\fr12)$ corresponding to the
  variables usually added to the Minkowski space (Lorentz vector) 
  variables to construct standard superspace. These could represent Dirac
  spinor degrees of freedom coupled to gravity.
  \end{itemize}
\item  spin $1$
  \begin{itemize}
  \item a Lorentz vector $(\fr12,\fr12)$. In standard superspace, the Minkowski
  space variables, these could correspond to Maxwell degrees of freedom.
  \item $(1,0)+(0,1)$ representations corresponding to self- and anti-self-dual
  halves of an antisymmetric $4\times 4$ matrix. The tangent vector fields 
  may, for certain specific examples, be chosen to be the generators of the 
  Lorentz group.
  \end{itemize}
\item spin $\fr32$
  \begin{itemize}
  \item  a Rarita-Schwinger representation $(1,\fr12)+(\fr12,1)$. The corresponding
  $X$'s and $Y$'s are the putative gravitino phase-space variables. 
  \item  a $(0,\fr32)+(\fr32,0)$ representation.
  \end{itemize}
\item spin $2$
  \begin{itemize}
  \item a $(1,1)$ representation corresponding to a
  symmetric tracefree four dimensional matrix  which together with the $(0,0)$
  representation corresponds to degrees of freedom transforming as the 
  gravitational metric $g_{\mu\nu}$.
  \item a $(\fr32,\fr12)+(\fr12,\fr32)$ representation.
  \item a $(0,2)+(2,0)$ Weyl tensor representation.
  \end{itemize}
\end{enumerate}
\goodbreak
Even for this rather restricted set of $15$ representations, the number of 
quadratic equations for the set of structure constants $t,u,c$ is rather 
formidable. For the choice of $\La_2=\La_2^{\sc{max}}$, 
the lattice $\O_2$ defined in \re{omega} has $339$ points,
yielding this number of $u$ structure constants.
The ordered lattice $\O_2\mid_{S\leq R}$ has $196$ points, in one-to-one
correspondence with the {\it a priori} $t$-structure constants, of which $33$ 
are zero in virtue of  \re{zeroes}. We therefore have 
{\it a priori} a total 
of $\ 163\, t\ +\ 339\, u\ +\ 15\, c\,=\, 517$ parameters, for which there are 
$1993$ $tt$-equations \re{ttequations}, $5569$ $tu$-equations \re{tuequations}
and $163$ $tuc$-equations \re{tcequations}, the latter being in one-to-one
correspondence with the number of $t$'s. It is remarkable that this highly 
over-determined set of $7725$ equations for a total of $517$ parameters has
any solutions at all. In fact the space of solutions is far from trivial; and
in spite of the phenomenal over-determination the structure of solutions is 
rather intricate. 
We have archived these equations in an electronic appendix at the URL
given in Appendix A.

\section{Examples}
A full discussion of all the allowed solutions would be rather involved and 
perhaps not entirely interesting. We restrict ourselves to a discussion 
of some classes of solutions of potential physical interest, 
based on subsets $\ \La_2 \subset \La_2^{\sc{max}}$.  
One simplifying restriction is to choose {\it a priori} the values of the $c$'s 
to be either $0$ or $1$. 
It is then convenient to label the points of $ \La_p $ by the corresponding
chosen values of $c$, writing
$\ \wh \La_p = \{\ (s,\ds)_{c(s,\ds)} \in (\bK\otimes\bK)_{\bZ_2} \}$.
We note that if any $c$ is chosen to be $0$, the corresponding $Y$
effectively decouples from $\cg$. In fact, this means that  
$X(0,1)+X(1,0)$ generating Lorentz transformations 
(with $c(0,1)=c(1,0)=0$) are always implicit.

\subsection{A purely bosonic example: 
$\wh\La_2 =\{ (0,0)_1\,,\,(1,1)_1\,,\,(\half,\half)_1\}$}  

This simple restriction to a configuration space containing variables
transforming like a gravitational metric and a Maxwell field allows solution
of the Jacobi identities in full generality. 
The $tt$ and $tuc$ equations imply that only two non-zero $t$'s
are allowed, namely, $t(0,0,1,1,1,1)$ and $t(0,0,\fr12,\fr12,\fr12,\fr12)$.
All $tt$ equations are then automatically satisfied. There are nine $u$ 
parameters which are possibly non-zero. We abbreviate them thus: 
\be \begin{array}{llll}
&u(0,0,0,0,0,0)\ =\ \uzm \,,
\quad &u(0,0,\fr12,\fr12,\fr12,\fr12)\ =\ \vmum \,,
  \quad &u(0,0,1,1,1,1)\ =\ \umum \,,          \\
&&u(\fr12,\fr12,0,0,\fr12,\fr12)\ =\ \vmdm\,,
 \quad &u(1,1,0,0,1,1)\ =\ \umdm \,,           \\
& u(\fr12,\fr12,1,1,\fr12,\fr12)\ =\ \wmdm \,,                 
  \quad &u(\fr12,\fr12,\fr12,\fr12,0,0)\ =\ \vmtm  \,,      
   \quad &u(1,1,1,1,0,0)\ =\ \umtm \,.    
\ea\la{ubosonic}\ee
The $tuc$ equations then fix the two non-zero $t$'s in terms of the $u$'s,
\be
t(0,0,1,1,1,1)
        =\umtm{-} \umum \quad,\quad
t(0,0,\fr12,\fr12,\fr12,\fr12)
        =\vmtm{-}\vmum\quad ,
\la{tbosonic}\ee
Only $tu$ equations remain. They are of two types, namely
\be
\umdm \umtm\ =\ \umdm \vmtm\ =\ \umdm \wmdm\ =\ \vmdm \umtm\ =\ 
\vmdm \vmtm\ =\ 0 
\label{cond2a}
\ee
and
\be\arr
\umdm(\uzm-2\umum)\,=\, 0\,,   &     \umtm(\uzm-\umtm)\,=\, 0\,,   & 
\\
\vmdm(\uzm-2\vmum)\,=\, 0\,,    &    \vmtm(\uzm-\vmtm)\,=\, 0\,,   &
\wmdm(\umum-2\vmum+\vmtm)\,=\, 0\,.  
\label{cond2b}
\ea\ee
The structure constants involving $X(0,0)$ (i.e. both $t$'s in \re{tbosonic} 
and the $u$'s in the first row of  \re{ubosonic}) merely determine the 
scaling properties of all the $X$'s and $Y$'s.
These constraints can be resolved in 12 independent ways. We list these
in \tableeleven\  of Appendix B.
For instance, case 12 on the table corresponds to the Lie algebra with
non-zero $t$ and $u$ structure constants,
\be\arr
&t(0,0,1,1,1,1)\, =\, \U_0-u_{-1}  \ ,
&t(0,0,\fr12,\fr12,\fr12,\fr12) \, =\, \fr12(\U_0-u_{-1}) \ , \\
&u(0,0,1,1,1,1) \, =\,u_{-1}\ ,
&u(0,0,\fr12,\fr12,\fr12,\fr12)\, =\, \fr12(u_{-1}{+}\U_0)\ , \\
&u(\fr12,\fr12,1,1,\fr12,\fr12)\, =\, \W_{-2}\ , \\
&u(0,0,0,0,0,0)\,=\, u(1,1,1,1,0,0)\, =
&u(\fr12,\fr12,\fr12,\fr12,0,0)\, =\, \U_{0} \ ,
\ea\ee
with free parameters $\U_0,\W_{-2}\in \bC\smallsetminus\{0\}$ and $u_{-1}\in\bC$.

\subsection{The simplest `super' example: 
$\wh\La_2 =\{ (\half,1)_1\,,\,(1,\half)_1\,,\,(0,0)_1\,,\,(1,1)_1\}$}
\la{second}
This restriction to the representations occurring in simple supergravity,
namely the graviton and gravitino representations, is of potential interest
for canonical formulations of simple $N{=}1$ supergravity theories.
Again, the complete set of solutions to the super Jacobi identities can
be found without any simplifying assumptions. The $tt,tu$ and $tuc$ equations
reduce the set of allowed structure constants to the following $u$'s, which are
possibly zero, \\[5pt]
{$
\begin{array}{l l l l}
u(0,0,0,0,0,0) =\uzm    &u(0,0,1,1,1,1) =\umum 
&u(1,1,0,0,1,1) =\umdm  &u(1,1,1,1,0,0) =\umtm      \\
u(0,0,\fr12,1,\fr12,1) = \unum
&u(\fr12,1,0,0,\fr12,1)= \undm
&u(\fr12,1,\fr12,1,0,0)= \untm
&u(\fr12,1,1,1,\fr12,1)= \unqm       \\
u(0,0,1,\fr12,1,\fr12)= \utum
&u(1,\fr12,0,0,1,\fr12)= \utdm
&u(1,\fr12,1,\fr12,0,0)= \uttm
&u(1,\fr12,1,1,1,\fr12)= \utqm       
\ea$}\\[5pt]
where the $u$'s with a non positive index are invariants under the $\bZ_2$ 
chiral transformation, $\, s\leftrightarrow\ds\,$, which relates the $u$'s
with positive indices thus: $\, u_i \leftrightarrow\wt u_i\,$. In virtue of the 
$tuc$-equations, the three non-trivial $t$'s are given in terms of the $u$'s by
\be
t(0,0,1,1,1,1)
        =\ u_{-3}{-} u_{-1}\ ,\quad
t(0,0,\fr12,1,\fr12,1)
        =\ u_3{-}u_1\ ,\quad
t(0,0,1,\fr12,1,\fr12)
        =\ \wt{u}_3{-}\wt{u}_1 \ .
\ee
The remaining $tu$ conditions are of two types, namely
\be\arr
&&\umdm \umtm\ =\ \umtm \undm \ =\ \umdm \untm \ =\ 
                          \umdm \unqm \ =\ \undm \untm \ =\ 0\,,
\\
&& \umdm \uttm \ =\ \umdm \utqm \ =\ \umtm \utdm \ =\ 
          \undm \uttm \ =\ \utdm \untm \ =\ \utdm \uttm \ =\ 0
\label{cond1a}
\ea\ee
and
\be\begin{array}{lll}
\undm(\uzm-2\unum)=0\,,  &\utdm(\uzm-2\utum)=0\,,  &\umdm(\uzm-2\umum)=0\,, 
\\
\untm(\uzm-\untm)=0\,,   &\uttm(\uzm-\uttm)=0\,,    &\umtm(\uzm-\umtm)=0\,,
\\
\unqm(\untm-2\unum+\umum)=0\,,   &\utqm(\uttm-2\utum+\umum)=0\,. 
\la{cond1b}\ea\ee
These constraints have  a set of $48$ independent solutions
given in the two parts of \tablefourtyeight\ in Appendix B.

Let us consider the case with the largest number of parameters not required to
be zero, namely the case $\a=5$ from \tablefourtyeight. It has non-trivial 
$t$ and $u$ structure constants,
\be\arr
&t(0,0,1,1,1,1)\,=\, \U_0-u_{-1}\ ,
&t(0,0,\fr12,1,\fr12,1)\,=\, \fr12(\U_0-u_{-1}) \ ,\\
&u(0,0,1,1,1,1)\,=\, u_{-1}\ ,
&u(0,0,\fr12,1,\fr12,1)\,=\, \fr12(u_{-1}{+}\U_0)\ , \\
&u(0,0,0,0,0,0)\, =\, u(\fr12,1,\fr12,1,0,0)\, =
&u(1,1,1,1,0,0) \, =\,  \U_{0} \\
&u(\fr12,1,1,1,\fr12,1) \, =\, \U_4\ 
\ea\ee
together with further non-zero structure constants obtained by interchanging 
dotted and undotted arguments of fermionic index pairs $S=(s,\ds)$, e.g. 
$t(0,0,\fr12,1,\fr12,1) \mapsto t(0,0,1,\fr12,1,\fr12)$. The free parameters 
are $\,\U_0\,,\U_4\in \bC\smallsetminus\{0\}\,$ and $\,u_{-1}\in\bC\,$.

\subsection{$\wh\La_2 =\{(\half,0)_1\,,\,(0,\half)_1\,,\,(\half,1)_1\,,
\,(1,\half)_1\,,\,(0,0)_1\,,\,(1,1)_1\}$}
\la{third}
The Rarita-Schwinger vector-spinor contains not only the gravitino 
representations $(\fr12,1)+(1,\fr12)$ included in \ref{second}, but also 
spinorial $(\fr12,0)+(0,\fr12)$ auxiliary (gauge) degrees of freedom. Including 
these significantly alters the structure of the solution space. In particular
the $tt$-equations have more solutions. These can be classified in a
straightforward albeit lengthy fashion as follows. 

The $tuc$-equations immediately imply that
\be\arr
&& t(\fr12,1,1,1,\fr12,1)=0\ , \quad  t(1,\fr12,1,1,1,\fr12)=0\ , \\[4pt]
&& u(1,1,\fr12,1,\fr12,1)=0\ , \quad  u(1,1,1,\fr12,1,\fr12)=0\,, \quad
u(1,1,1,1,1,1)=0\ .
\ea\ee
The $tuc$-equations are then entirely resolved and the
discussion of the $tt$-equations may be carried out most conveniently 
in terms of the following parameters, which may possibly take zero value,
\be\arr
&&t(0,\fr12,1,\fr12,1,1)\ = \   \tnum  \quad,\quad
t(0,\fr12,1,1,1,\fr12)\ = \   \tndm   \quad,\quad
t(\fr12,1,1,1,\fr12,0)\ = \   \tntm   \quad, \\[4pt]
&&t(\fr12,0,\fr12,1,1,1)\ = \   \ttum    \quad,\quad
 t(\fr12,0,1,1,\fr12,1)\ = \   \ttdm  \quad,\quad
 t(1,\fr12,1,1,0,\fr12)\ = \   \tttm \quad.
\ea\ee
The $tt$-equations yield the following $10$ quadratic constraints among
$t_i$ and $\wt t_i, i=1,2,3$:
\be 
t_i\ \wt t_j = 0\ ,\ i\neq j \ ,\quad  
t_1\  t_j  =0\  ,\ j \neq 1  \ ,\quad
\wt t_1\  \wt t_j =0\ ,\ j \neq 1\ .
\la{cond3}\ee 
Introducing the shorthand for the remaining $t$-parameters,
\be\arr
&&t(0,0,\fr12,1,\fr12,1)\ = \   \wnum \quad,\quad
t(0,0,1,\fr12,1,\fr12)\ = \   \wtum \quad,\\[4pt]
&&t(0,0,1,1,1,1)\ = \   \tmum \quad,\quad
t(0,0,0,\fr12,0,\fr12)\ = \   \vnum \quad,\quad
t(0,0,\fr12,0,\fr12,0)\ = \   \vtum \quad  ,
\ea\ee
we obtain twelve disjoint solutions of the $tt$-equations which we present
in tabulated form: \tabletwelve of Appendix B. 
Every line of the table is a solution of the entire set of $tt$-equations. 

Our classification of all solutions of the $tu$-equations is too lengthy to
include here. A file containing this may be obtained from the authors by e-mail.
Here, we concentrate on one case which seems particularly interesting:
Case 4 on \tabletwelve. This has $\bZ_2$  chiral symmetry under 
$\,s \leftrightarrow \ds\,$ and the 
property that $X(1,1)$ commutes with all fermionic $X$'s, namely
\be\arr
&t(0,\fr12,1,\fr12,1,1)\ = \   \tnum\  \neq\ 0  \ ,\qquad
t(\fr12,0,\fr12,1,1,1)\ = \   \ttum \  \neq\ 0  \ , \\[4pt]
&t(0,\fr12,1,1,1,\fr12)\ = \  
t(\fr12,0,1,1,\fr12,1)\ = \  
t(\fr12,1,1,1,\fr12,0)\ = \  
t(1,\fr12,1,1,0,\fr12)\ = \ 0 \ .
\ea\ee
The $tu,tuc$ equations then imply that the only possibly non-zero $u$'s
remaining are:
\be\arr
& u(0,0,0,0,0,0)\ =\ u(0,0,1,1,1,1)\ =\ u_0   \\[3pt]
& u(0,0,0,\fr12,0,\fr12) \ =\ 
 u(0,0,\fr12,0,\fr12,0) \ =\ \fr{u_0}2\\[3pt]
& u(0,0,\fr12,1,\fr12,1)\ =\
 u(0,0,1,\fr12,1,\fr12) \ =\  \fr{u_0}2\\[3pt]
& u(0,\fr12,1,1,1,\fr12) \ =\  t_1\ ,\quad
u(\fr12,0,1,1,\fr12,1) \ =\  \wt t_1 \\[3pt]
&u(\fr12,1,0,0,\fr12,1) \ =\  u_2\ ,\quad 
u(1,\fr12,0,0,1,\fr12) \ =\  \wt u_2  \\[3pt] 
&u(\fr12,1,1,1,\fr12,1) \ =\  u_4 \ ,\quad 
u(1,\fr12,1,1,1,\fr12) \ =\  \wt u_4\\[3pt]
&u(\fr12,1,\fr12,1,0,0) \ =\  
u(1,\fr12,1,\fr12,0,0) \ =\ u(1,1,1,1,0,0) = u_3
\ea\ee
All $tt,tuc$-equations are then solved and the remaining $tu$-equations
may be solved in two possible ways:

a) $u_3 =0\ $,

b) $u_3 =u_0\ ,\quad u_2 = \wt u_2 = u_4 = \wt u_4\ =\ 0\ $.

\noindent
The latter, for instance, corresponds to the simple superalgebra with non-zero
$t$ and $u$ structure constants,
\be\arr
 & t(0,0,0,\fr12,0,\fr12)\, =\, - \fr12 u_0\ ,    
 & t(0,0,\fr12,1,\fr12,1)\, =\, \fr12 u_0\ ,\\[3pt]
& t(0,\fr12,1,\fr12,1,1) \, =\,  t_1\ ,
& u(0,\fr12,1,1,1,\fr12)\, =\, t_1\ ,\\[3pt]
& u(0,0,0,\fr12,0,\fr12)\, =\,\fr12 u_0\ , 
& u(0,0,1,\fr12,1,\fr12)\, =\, \fr12 u_0\ , \\[3pt]
&u(1,\fr12,1,\fr12,0,0)\, =\, u(1,1,1,1,0,0)\, =
& u(0,0,0,0,0,0)\, =\, u_0\ ,\\[3pt]
& u(0,0,1,1,1,1) \, =\, u_0
\ea\ee
and further structure constants obtained by interchanging dotted and undotted
arguments of fermionic index pairs together with the replacement of
$t_1$ by $\wt t_1$. Here $t_1,\wt t_1$ are non-zero free parameters and
$u_0$ is arbitrary.

\subsection{Adding a vector: \hfill\break
$\wh\La_2 =\{(\half,\half)_1\,,\,(\half,0)_1\,,\,(0,\half)_1\,,
\,(\half,1)_1\,,\,(1,\half)_1\,,\,(0,0)_1\,,\,(1,1)_1\}$}
\la{first}
Adding a vector, corresponding to a Maxwell degree of freedom, to the set of 
representations in \ref{third}, yields an example
which has the additional interesting feature of combining the super-Poincar\'e
representations in (\ref{st1},\ref{st2})  with the simple supergravity 
representations in \ref{second}. 
The super-Jacobi identities immediately imply that
\be
t(\fr12,\fr12,1,1,\fr12,\fr12) = 0 \quad,\quad 
t(1,\fr12,1,1,1,\fr12) = 0 \  .
\ee
The complete discussion of all the solutions of the super-Jacobi identities
is rather detailed. The assumption of the {\it super-Poincar\'e condition},
\be
t(0,\fr12,\fr12,0,\fr12,\fr12)\neq 0\ ,
\ee
considerably simplifies further discussion. 
In particular, apart from the six parameters of the form $t(0,0,a,b,a,b)$
encoding the scaling behaviours of the $X$'s, only six further $t$'s remain 
non-zero. The $tt,tu,tuc$-equations yield a solution with $19$ parameters,
$\{\, u_0\,,u_i\,,\wt u_i\,; i=1,\dots,9\}$ invariant under the $\bZ_2$
{\it chirality transformation},
\be 
  u_i \leftrightarrow \wt u_i \quad,\quad s\leftrightarrow \ds \quad.
\la{chirality}
\ee
Modulo this symmetry, the non-zero $t$ and $u$ structure constants are
\be\arr
&t(0,\fr12,\fr12,0,\fr12,\fr12)\,=\, (u_2{+}\wt u_2)\ ,
&t(0,\fr12,\fr12,1,\fr12,\fr12)\,=\, (u_6{-}u_3)\ ,\\[3pt]
&t(0,\fr12,1,\fr12,1,1)\,=\, (u_4{+}\wt u_8)\ ,
&t(\fr12,1,1,\fr12,\fr12,\fr12)\,=\, (u_7{+}\wt u_7)\ ,\\[3pt]
&u(0,\fr12,\fr12,\fr12,\fr12,0)\,=\, u_2\ ,\ 
u(0,\fr12,\fr12,\fr12,\fr12,1) \,=\,  u_3\ ,
&u(0,\fr12,1,1,1,\fr12) \,=\, u_4\ ,\\[3pt] 
&u(\fr12,1,\fr12,\fr12,0,\fr12) \,=\, u_6\ ,\
u(\fr12,1,\fr12,\fr12,1,\fr12) \,=\, u_7\ ,
&u(\fr12,1,0,0,\fr12,1) \,=\, u_5\ ,\\[3pt]
&u(\fr12,1,1,1,\fr12,0) \,=\, u_8\ ,\ \,
u(\fr12,1,1,1,\fr12,1) \,=\, u_9\ ,
&u(0,\fr12,0,0,0,\fr12) \,=\, u_1\ ,
\ea\ee
together with the scaling rules,
\be\arr
&t(0,0,0,\fr12,0,\fr12) \,=\, -\fr{u_0}{2}\ ,
&t(0,0,1,\fr12,1,\fr12) \,=\, -\fr{u_0}{2}\ , \\[3pt]
&t(0,0,\fr12,\fr12,\fr12,\fr12)\,=\, -u_0\ ,
&t(0,0,1,1,1,1) \,=\, -u_0\ ,\\[3pt]
&u(0,0,0,\fr12,0,\fr12) \,=\,  \fr{u_0}{2}\ ,
&u(0,0,1,\fr12,1,\fr12) \,=\,  \fr{u_0}{2}\ , \\[3pt]
&u(0,0,\fr12,\fr12\fr12,\fr12) \,=\,  u_0\ ,
&u(0,0,1,1,1,1) \,=\,  u_0\ ,\\[3pt]
&u(0,0,0,0,0,0) \,=\, u_0\ .
\ea\ee

\subsection{An extension of the \sp: 
$\La_2 = \{(s,\ds)\,;\  0 < s{+}\ds \leq 2\}$}
In \cite{DN2} we constructed an explicit example of a spin $\fr32$ superalgebra
with the \sp\ as a subalgebra. Following the procedure of section 4.3 of 
\cite{DN2}, we may extend that example to include elements of spin $2$ as well,
maintaining the super Poincar\'e embedding. This example has every 
representation with spin $\leq 2$, except for the scalar $(0,0)$. The 
identification of $X(0,1)+X(1,0)$ with the Lorentz generators leads to the
decoupling of $Y(0,1)+Y(1,0)$, which may be put to zero with $c(0,1)=c(1,0)=0$.
All other $12$ $c$'s are taken to be $1$. 

There are $46$ free parameters, which we denote as follows:
\be\begin{array}{rllll}
u(\fr12,0\,,\fr12,\fr12\,,0,\fr12)= u_1\,,&& 
u(1,\fr12\,,\fr12,\fr12\,,\fr12,0)= u_2\,,&&
u(\fr12,0\,,\fr12,\fr12\,,1,\fr12)= u_3\,,\\[3pt]
u(1,\fr12\,,\fr12,\fr12\,,\3h ,0)=u_4\,,&&
u(\3h,0\,,\fr12,\fr12\,,1,\fr12)= u_5\,,&&
u(1,\fr12\,,\fr12,\fr12\,,\fr12,1)=u_6\,,\\[3pt]
u(\fr12,0\,,\fr12,\3h\,,0,\3h)= u_7\,,&& 
u(\fr12,0\,,1,1\,,\fr12,1)= u_8\,,&&
u(\fr12,0\,,\3h,\fr12\,,1,\fr12)= u_9\,,\\[3pt]
u(\fr12,0\,,2,0\,,\3h ,0)=u_{10}\,,&&
u(1,\fr12\,,\fr12,\3h\,,\fr12,1)=u_{11}\,,&&
u(1,\fr12\,,1,1\,,0,\fr12)= u_{12}\,,\\[3pt]
u(1,\fr12\,,1,1\,,0,\3h)= u_{13}\,,&&
u(1,\fr12\,,1,1\,,1,\fr12)= u_{14}\,,&&
u(1,\fr12\,,\3h,\fr12\,,\fr12,0)= u_{15}\,,\\[3pt]
u(1,\fr12\,,\3h,\fr12\,,\fr12,1)=u_{16}\,,&&
u(1,\fr12\,,\3h,\fr12\,,\3h ,0)=u_{17}\,,&&
u(1,\fr12\,,2,0\,,1,\fr12)= u_{18}\,,\\[3pt]
u(\3h,0\,,1,1\,,\fr12,1)=u_{19}\,,&&
u(\3h,0\,,\3h,\fr12\,,0,\fr12)= u_{20}\,,&&
u(\3h,0\,,\3h,\fr12\,,1,\fr12)= u_{21},  \\[3pt]
u(\3h,0\,,2,0\,,\fr12,0)= u_{22}\,,&&
u(\3h,0\,,2,0\,,\3h ,0)=u_{23}\,,&&
\la{5u}\ea\ee
together with further $23$ parameters obtained from the above under the 
transformation \re{chirality}.
All other $u$'s are zero, except for those concerning the Lorentz generators
$X(0,1)+X(1,0)$, viz. $u(0,1\,,a,b\,,a,b)$ and $u(1,0\,,a,b\,,a,b)$, which 
together with the `Lorentz' $t$'s (those containing $(0,1)$ or $(1,0)$ in any 
of the three positions) are completely determined by Lorentz covariance. 
The remaining non-zero $t$'s take the form,
\be\begin{array}{rllllll}
&t(0,\fr12\,,\fr12,0\,,\fr12,\fr12) &=& u_1\,+\, \wt u_1\ ,
&t(\fr12,0\,,1,\fr12\,,\fr12,\fr12) &=& u_2\,-\, u_3\ , \\[4pt]
&t(\fr12,1\,,1,\fr12\,,\fr12,\fr12) &=& u_6\,+\, \wt u_6\ ,
&t(1,\fr12\,,\3h,0\,,\fr12,\fr12) &=& u_4\,+\, u_5\ , \\[4pt]
&t(\fr12,0\,,0,\3h\,,\fr12,\3h) &=& u_7\,+\, \wt u_{20}\ ,
&t(\fr12,0\,,\fr12,1\,,1,1) &=&  u_8\,+\, \wt u_{12}\ , \\[4pt]
&t(\fr12,0\,,1,\fr12\,,\3h,\fr12) &=&  u_9\,+\,  u_{15}\ ,
&t(\fr12,0\,,\3h,0\,,2,0) &=&  u_{10}\,+\,  u_{22}\ , \\[4pt]
&t(\fr12,1\,,1,\fr12\,,\fr12,\3h) &=&  u_{11}\,-\,  \wt u_{16}\ ,
&t(0,\3h\,,1,\fr12\,,1,1) &=&  u_{13}\,-\,  \wt u_{19}\ , \\[4pt]
&t(1,\fr12\,,\3h,0\,,\3h,\fr12) &=& u_{21}\,-\,   u_{17}\ ,
\la{5t}\ea\ee
together with $9$ further $t$'s obtained under the transformation \re{chirality}
and using the relations \re{skewsym} to recover $t$'s in the ordered set. 
With the above structure constants, all the $tt,tu,tuc$-equations are
resolved. The non-zero supercommutation relations may be read off directly 
from \re{5u} and \re{5t}. Setting the parameters 
$\{ u_i, \wt u_i\ ; i=7,\dots,23\}$ to zero may easily be seen to reduce this
superalgebra to the $12$-parameter spin $\fr32$ extension of the \sp\  
obtained in \cite{DN2}. Moreover, coordinate representations of this as well
as the previous examples in this section may be found following that reference.

\section{Concluding remarks}
We have developed a framework for the construction and investigation of 
Lorentz covariant Heisenberg superalgebras with generators transforming
according to representations of arbitrary values of spin. We have thus obtained
a complete parametrisation of all such superalgebras, for the case of unit 
multiplicity. The parameter space is highly overdetermined. Closer 
investigation, however, reveals surprisingly non-trivial possibilities of 
resolving the constraints. As an
example, we have obtained the most general set of constraints for superalgebras
containing generators of spins up to two; and we have found several classes of
explicit solutions. We have remained in a broader algebraic setting. 
Our spin two superalgebras, however, have the algebraic structure of phase 
spaces possibly underlying gravity and supergravity models.
Concrete physical application of our algebras, for instance, to the canonical 
quantisation of supergravity theories, remains for future investigation.

\noindent
{\bf Acknowledgments} 

\noindent
One of us (J.N.) presented this work at the first annual meeting of the 
TMR European Network ``Integrability, non-perturbative effects and symmetry in 
quantum field theory'' (contract number FMRX-CT96-0012) held at 
Santiago de Compostela in September 1998. The other (C.D.) acknowledges the 
hospitality of the Albert-Einstein-Institut, Potsdam, a node of this network,
where part of this work was performed.

\goodbreak
\appendix
\baselineskip=14pt
\section{The spin two structure equations}
We have archived the complete set of structure equations for spin two 
superalgebras at the URL\\
{\cmssl http://www.mis.mpg.de/preprints/98/preprint1598-addendum1.html}\\
The files given there list a) the $7725$ quadratic polynomials which
need to vanish for the satisfaction of the super Jacobi identities; and 
b) the 196 t's and 339 u's, which together with the 15 c's yield the 
total of 517 constrained parameters. The equations given are labeled as
follows
$$
\begin{array}{lll}
\mbox{the tuc-equations:} &tcrl(i);& i=1,...,163 \\[4pt]
\mbox{the tt-equations:}  &ttrl(i);& i=1,...,1993\\[4pt]
\mbox{the tu-equations:}  &tu0rl(i);& i=1,...,324 \\[4pt]
                       &tu1rl(i);& i=1,...,1020 \\[4pt]
                       &tu2rl(i);& i=1,...,1688 \\[4pt]
                       &tu3rl(i);& i=1,...,1694 \\[4pt]
                       &tu4rl(i);& i=1,...,843 
\ea
$$ 
Here  $\ tuNrl\ $ with $\ N=0,\dots,4\ $  denote relations derived from 
super Jacobi identities involving $X(s,\ds)$, $X(r,\dr)$ and $Y(v,\dv)$ for 
$S\leq R$ and with $\ N=2(s+\ds)$. The equations are given in a text format
compatible with {\small REDUCE} or {\small MAPLE}.

\section{Solutions of the spin two structure equations}
In this appendix we give details of the solutions discussed in section 4.
The following table lists the $12$ independent solutions of \re{cond2a} and 
\re{cond2b} ($\wh\La_2 =\{ (0,0)_1\,,\,(1,1)_1\,,\,(\half,\half)_1\}$). 
\vskip 0.2 true cm
\noindent
{\cmssll{\tableeleven}}\\[2pt]
\noindent
{\small
\begin{tabular}
{|c  |c   |c   |c   |c   |c   |c   |c   |c   |c   |c   |c   |c   | }
\hline
  &\uz   &\umu  &\umd  &\umt  &\vmu  &\vmd  &\vmt  &\wmd     \\
\hline\hline
1 &\uz    &\umu  &0     &0     &\vmu    &0      &0     &0        \\
\cline{1-1}\cline{3-4}
2 &     &\uz/2 &\Umd  &      &        &       &      &         \\
\cline{1-1}\cline{3-4}\cline{9-9}
3 &       &2\vmu&0     &      &          &      &      &\Wmd      \\
\cline{1-1}\cline{3-3}\cline{6-7}
4 &       &\uz  &      &    &\uz/2       &\Vmd   &      &         \\
\cline{1-1}\cline{3-3}\cline{9-9}
5 &       &\umu &      &    &            &      &      &0        \\
\cline{1-1}\cline{3-4}
6 &       &\uz/2&\Umd   &    &            &      &      &         \\
\cline{1-4}\cline{6-8}
7 &\Uz    &\umu &0     &    &\vmu        &0     &\Uz   &         \\
\cline{1-1}\cline{6-6}\cline{9-9}
8 &       &     &      &    &(\umu+\Uz)/2&      &      &\Wmd      \\
\cline{1-1}\cline{5-6}\cline{8-9}
9 &       &     &      &\Uz &\vmu        &      &0     &0        \\
\cline{1-1}\cline{6-6}\cline{9-9}
10 &       &     &      &    &\umu/2      &      &      &\Wmd      \\
\cline{1-1}\cline{6-6}\cline{8-9}
11&       &     &      &    &\vmu        &      &\Uz   &0        \\
\cline{1-1}\cline{6-6}\cline{9-9}
12&       &     &      &    &(\umu+\Uz)/2&      &      &\Wmd      \\
\hline\hline
\end{tabular}
}
\vskip 0.2 true cm\noindent
The entries give the parameters listed in \re{ubosonic} in terms of
the free subset. If the $u_{x}$ column is marked $u_{x}$ it means that it is a 
completely free parameter. If it is marked in uppercase roman font, e.g., 
{\sc U}$_{x}$, it means that it is a free 
parameter but necessarily non-zero. Also in this notation, the $48$ distinct
choices of unconstrained parameters satisfying \re{cond1a} and  \re{cond1b}
(section 4.2, 
$\ \wh\La_2 =\{ (\half,1)_1\,,\,(1,\half)_1\,,\,(0,0)_1\,,\,(1,1)_1\}$)  
are listed in \tablefourtyeight.
\goodbreak

\noindent
{\cmssll{\tablefourtyeight}}\\[2pt]
{\footnotesize
\begin{tabular}
{|c  |c   |c   |c   |c   |c   |c   |c   |c   |c   |c   |c   |c   | }
\hline
$\alpha$
  &\uz  &\umu  &\umd &\umt&\unu  &\utu  &\und &\utd &\unt&\utt&\unq&\utq\\
\hline\hline
1 &\uz  &\uz/2 &\Umd  &0   &\uz/2 &\uz/2 &\Und  &\Utd  &0   &0   &0   &0 \\
\cline{1-1}\cline{7-7}\cline{9-9}
2 &     &      &     &    &      &\utu   &   &0    &    &    &    &   \\
\cline{1-1}\cline{6-9}
3 &     &      &     &    &\unu  &\uz/2 &0    &\Utd  &    &    &    &   \\
\cline{1-1}\cline{7-7}\cline{9-9}
4 &     &      &     &    &      &\utu  &     &0    &    &    &    &   \\
\cline{1-7}\cline{10-13}
5 &\Uz  &\umu  &0 &\Uz &(\Uz+\umu)/2&(\Uz+\umu)/2 & &  &\Uz&\Uz&\Unq&\Utq \\
\cline{1-1}\cline{7-7}\cline{13-13}
6 &     &      &  &    &            &\utu         & &  &   &   &   &0   \\
\cline{1-1}\cline{6-7}\cline{12-13}
7 &     &      &  &    &\unu        &(\Uz+\umu)/2 & &  &  &  &0  &\Utq \\
\cline{1-1}\cline{6-7}\cline{11-12}
8 &     &      &  &    &(\Uz+\umu)/2&\umu/2        &  &   &  &0  &\Unq&  \\
\cline{1-1}\cline{6-7}\cline{10-11}
9 &     &      &  &    &\umu/2      &(\Uz+\umu)/2  & &  &0   &\Uz&  &  \\
\cline{1-1}\cline{6-7}\cline{10-10}\cline{12-13}
10&     &      &  &    &\unu        &\utu          & &  &\Uz &  &0  &0   \\
\cline{1-1}\cline{6-6}\cline{11-12}
11&     &      &  &    &(\Uz+\umu)/2&              & &  &   &0  &\Unq&    \\
\cline{1-1}\cline{6-6}\cline{10-11}
12&     &      &  &    &\umu/2      &              & &  &0  &\Uz&  &    \\
\cline{1-1}\cline{6-7}\cline{10-13}
13&     &      &  &    &\unu        &\umu/2      & &  &\Uz&0  &0  &\Utq \\
\cline{1-1}\cline{7-7}\cline{10-11}
14&     &      &  &    &            &(\Uz+\umu)/2 & &  &0  &\Uz&   &   \\
\cline{1-1}\cline{6-7}\cline{11-12}
15&     &      &  &    &\umu/2      &\umu/2      & &  &   &0  &\Unq&   \\
\cline{1-1}\cline{6-7}\cline{10-13}
16&     &      &  &    &\unu        &\utu        & &  &\Uz&0  &0  &0   \\
\cline{1-1}\cline{10-11}
17&     &      &  &    &            &            & &  &0  &\Uz&   &   \\
\cline{1-1}\cline{6-6}\cline{11-12}
18&     &      &  &    &\umu/2      &            & &  &   &0  &\Unq&   \\
\cline{1-1}\cline{6-7}\cline{12-13}
19&     &      &  &    &\unu        &\umu/2      & &  &   &   &0  &\Utq  \\
\cline{1-1}\cline{7-7}\cline{13-13}
20&     &      &  &    &            &\utu        & &  &   &   &   &0    \\
\cline{1-3}\cline{5-9}\cline{12-13}
21&\uz  &\uz  &  &0 &\uz/2&\uz/2&\Und&\Utd         &  &   &\Unq&\Utq  \\
\cline{1-1}\cline{13-13}
22&     &     &  &  &     &     &   &            & &   &  &0    \\
\cline{1-1}\cline{12-13}   
23&     &     &  &  &     &     &   &            & &   &0  &\Utq  \\
\cline{1-1}\cline{3-3}\cline{13-13}
24&     &\umu &  &  &     &     &   &            & &   &   &0    \\
\cline{1-2}\cline{6-13}
25&\Uz  &     &  &  &(\Uz+\umu)/2&(\Uz+\umu)/2&0&0 &\Uz&\Uz&\Unq&\Utq \\
\cline{1-1}\cline{7-7}\cline{13-13}
26&     &     &  &  &            &\utu        & &  &   &   &  &0   \\
\cline{1-1}\cline{6-7}\cline{12-13}
27&     &     &  &  &\unu       &(\Uz+\umu)/2 & &  &   &   &0  &\Utq \\
\cline{1-1}\cline{7-7}\cline{13-13}
28&     &     &  &  &           &\utu         & &  &   &   &   &0   \\
\cline{1-3}\cline{6-8}\cline{10-13}
29&\uz  &\uz  &  &  &\uz/2 &\uz/2  &\Und  &   &0   &0   &\Unq&\Utq \\
\cline{1-1}\cline{7-7}\cline{13-13}
30&     &    &  &  &      &\utu   &     &   &    &    &  &0   \\
\cline{1-1}\cline{3-3}\cline{12-13}
31&     &2\utu&  &  &      &       &     &   &    &    &0  &\Utq \\
\cline{1-1}\cline{3-3}\cline{13-13}
32&     &\umu &  &  &      &       &     &   &    &    &   &0   \\
\cline{1-1}\cline{3-3}\cline{6-9}\cline{12-13}
33&     &\uz  &  &  &\uz/2 &\uz/2  &0    &\Utd&    &    &\Unq&\Utq \\
\cline{1-1}\cline{3-3}\cline{6-6}\cline{13-13}
34&     &2\unu&  &  &\unu   &       &     &   &    &    &  &0   \\
\cline{1-1}\cline{3-3}\cline{12-13}
35&     &\uz  &  &  &      &       &     &   &    &    &0  &\Utq \\
\cline{1-1}\cline{3-3}\cline{13-13}
36&     &\umu &  &  &      &       &     &   &    &    &   &0   \\
\cline{1-2}\cline{6-7}\cline{9-10}\cline{12-13}
37&\Uz  &  &  &  &(\Uz+\umu)/2&\umu/2     & &0 &\Uz& &\Unq &\Utq \\
\cline{1-1}\cline{7-7}\cline{13-13}
38&     &     &  &  &            &\utu       & &  & &   &   &0   \\
\cline{1-1}\cline{6-7}\cline{12-13}
39&     &     &  &  &\unu        &\umu/2     & &  & &   &0   &\Utq \\
\cline{1-1}\cline{7-7}\cline{13-13}
40&     &     &  &  &            &\utu       & &  & &   &    &0 \\
\cline{1-1}\cline{6-7}\cline{10-13}
41&     &     &  &  &\umu/2     &(\Uz+\umu)/2& &  &0&\Uz&\Unq &\Utq \\
\cline{1-1}\cline{7-7}\cline{13-13}
42&     &     &  &  &           &\utu        & &  & &   &    &0 \\
\cline{1-1}\cline{6-7}\cline{12-13}
43&     &     &  &  &\unu       &(\Uz+\umu)/2& &  & &   &0   &\Utq \\
\cline{1-1}\cline{7-7}\cline{13-13}
44&     &     &  &  &           &\utu        & &  & &   &    &0 \\
\cline{1-2}\cline{6-7}\cline{11-13}
45&\uz  &     &  &  &\umu/2     &\umu/2      & &  & &0  &\Unq &\Utq \\
\cline{1-1}\cline{7-7}\cline{13-13}
46&     &     &  &  &           &\utu        & &  & &   &    &0 \\
\cline{1-1}\cline{6-7}\cline{12-13}
47&     &     &  &  &\unu       &\umu/2      & &  & &   &0   &\Utq \\
\cline{1-1}\cline{7-7}\cline{13-13}
48&     &     &  &  &           &\utu        & &  & &   &    &0 \\
\hline\hline
\end{tabular}
}

\noindent
For the examples of section 4.3 
($\wh\La_2 =\{(\half,0)_1\,,\,(0,\half)_1\,,\,(\half,1)_1\,,
\,(1,\half)_1\,,\,(0,0)_1\,,\,(1,1)_1\}$) the constraints \re{cond3} 
have the following $12$ possible choices of free sets of $t$'s.

\noindent
{\cmssll{\tabletwelve}}
\vskip 2pt
\noindent
{\small
\begin{tabular}
{|c |c   |c   |c   |c   |c   |c   |c   |c   | }
\hline
  &\tnu &\ttu  &\tnd  &\ttd  &\tnt  &\ttt  &\wnu      &\wtu  \\
\hline\hline
1  &0      &0     &\Tnd   &\Ttd   &0     &0     &\tmu$+$\vtu &\tmu$+$\vnu  \\
\cline{1-1}\cline{5-6}\cline{8-8}
2  &       &      &    &0         &\Tnt   &      &\vtu$-$\tmu &    \\
\cline{1-1}\cline{6-6}\cline{8-8}
3  &       &      &    &          &0     &      &\wnu      &    \\
\cline{1-4}\cline{8-8}
4  &\Tnu    &\Ttu   &0     &      &      &      &\tmu$-$\vtu &    \\
\cline{1-1}\cline{3-3}\cline{8-8}
5  &      &0     &      &          &      &      &\wnu      &    \\
\cline{1-2}\cline{5-5}\cline{7-9}
6  &0      &      &      &\Ttd   &      &\Ttt   &\vtu$+$\tmu &\vnu$-$\tmu  \\
\cline{1-1}\cline{7-9}
7  &       &      &      &       &      &0     &\vnu$+$\tmu &\wtu  \\
\cline{1-1}\cline{3-3}\cline{5-5}\cline{8-8}
8  &       &\Ttu   &      &0     &      &      &\tmu$-$\vnu &    \\
\cline{1-1}\cline{3-3}\cline{6-9}
9  &       &0     &      &      &\Tnt   &\Ttt   &\vtu$-$\tmu &\vnu$-$\tmu  \\
\cline{1-1}\cline{6-6}\cline{8-8}
10 &       &      &      &      &0     &      &\wnu      &    \\
\cline{1-1}\cline{6-9}
11 &       &      &      &      &\Tnt   &0     &\vtu$-$\tmu &\wtu  \\
\cline{1-1}\cline{6-6}\cline{8-8}
12 &       &      &      &      &0     &      &\wnu      &    \\
\hline\hline
\end{tabular}
}
\vskip 0.5 true cm
\noindent
We note that some of theses cases are related by the chiral $\bZ_2$ symmetry~:
$(2\leftrightarrow 6)$,
$(3\leftrightarrow 7)$, $(5\leftrightarrow 8)$ and
$(10\leftrightarrow 11)$.


\goodbreak
\end{document}